\documentclass[twoside]{article}
\usepackage{fleqn,espcrc2}
\usepackage{epsfig}
\usepackage{psfig}

\usepackage{graphicx}
\newcommand{\AmS}{{\protect\the\textfont2
  A\kern-.1667em\lower.5ex\hbox{M}\kern-.125emS}}

\title{Violation of Equivalence Principle and Solar Neutrinos
\thanks{Talk presented by H. Nunokawa at 
Europhysics Neutrino Oscillation Workshop (NOW2000), 
Otranto, Italy, September 9-16, 2000
}}

\author{A. M. Gago\address{
Secci\'{o}n F\'{\i}sica, Departamento de Ciencias,  
Pontificia Universidad Cat\'olica del Per\'u \\
Apartado 1761, Lima, Per\'u}$^{,}$\address{
Instituto de F\'{\i}sica,  Universidade de S\~ao Paulo 
C.\ P.\ 66.318, 05389-970 S\~ao Paulo, Brazil}%
        , H. Nunokawa\address{Instituto de F\' {\i}sica Gleb Wataghin, 
        Universidade Estadual de Campinas, UNICAMP \\    
     13083-970 Campinas SP, Brazil}
and          R. Zukanovich Funchal$^{\mbox{\scriptsize  b}}$}
\begin{document}

\begin{abstract}
We have updated the analysis for the solution to the solar neutrino 
problem by the long-wavelength neutrino oscillations induced 
by a tiny breakdown of the weak equivalence principle of 
general relativity, and obtained a very good fit 
to all the solar neutrino data.
\vspace{1pc}
\end{abstract}

\maketitle

\section{INTRODUCTION}

It is considered that the results coming from 
atmospheric neutrino experiments~\cite{atmospheric}, 
as well as solar neutrino 
experiments~\cite{homestake,sage,gallex,gno,sk00}, 
are strong evidence of neutrino 
oscillation indicating the presence neutrino mass 
and flavor mixing. 
However, several alternative explanations to 
these results, 
which do not invoke neutrino mass and/or mixing, 
exist and are not yet excluded~\cite{others}. 
%the dynamics underlying such conversion is yet to 
%be established by the future experiments. 
%

The interesting idea that gravitational forces may induce 
neutrino mixing and flavour oscillations, if the weak 
equivalence principle of general relativity is violated, 
was proposed  about a decade ago~\cite{gasper,hl}, 
and thereafter, many works have been performed on this
subject~\cite{muitas}. 

Many authors have investigated the possibility of solving the solar
neutrino problem (SNP) by such gravitationally induced  neutrino 
oscillations~\cite{vep_msw}, generally finding it necessary,
in this context, to invoke the MSW like resonance~\cite{hl} 
in order reduce appropriately the solar neutrino fluxes 
with the specific energy dependence, required to explain the data. 
Nevertheless we have demonstrated in Ref.~\cite{us} that 
all the recent solar neutrino data coming from gallium, 
chlorine and water Cherenkov detectors can be well accounted 
for also by long-wavelength neutrino oscillations induced 
by a violation of the equivalence principle (VEP).
In this talk, we present updated fit of such 
gravitationally induced long-wavelength oscillation 
solution~\cite{us,CDM} to the most recent solar neutrino data
which include the first GNO measurement~\cite{gno}. 
\vglue -1.0cm
\section{THE VEP FRAMEWORK}
We follow the framework proposed in Refs.~\cite{gasper,hl}. 
In the presence of violation of equivalence principle, 
neutrino mixing and oscillation can occur even if 
neutrinos are massless. 
In this work we assume oscillations only between 
two species of neutrinos, which are massless (or degenerate 
in mass), 
either between active and active ($\nu_e \leftrightarrow
\nu_\mu,\nu_\tau$) or  active and sterile
($\nu_e \leftrightarrow \nu_s$, $\nu_s$ being 
an electroweak singlet) neutrinos. 

To describe the VEP induced massless neutrino oscillation 
mechanism, phenomenologically, we can simply do 
the following replacement in the usual mass induced 
oscillation formula:
$\Delta m^2/2E \to 2E|\phi \Delta \gamma|$ 
and $\theta  \to \theta_G$, 
where 
$\Delta m^2$ is the mass squared difference, 
$\phi$ is the gravitational potential which is assume to be
constant in our work, $\theta$ is the usual mixing angle 
which relate weak and mass eigenstates 
and $\theta_G$ is the mixing which 
relates weak and gravitational eigenstates, 
and  $\Delta \gamma$ is the quantity which measures 
the magnitude of VEP. 
% , the difference of the gravitational couplings between the two 
% neutrinos involved normalized by the sum.

The survival probability of $ \nu_e$ produced in the Sun 
after traveling the distance $L$ to the Earth is given by, 
\begin{equation}
 P( \nu_e \rightarrow \nu_e) 
= 1 - \sin^2 2\theta_G \sin^2 \frac{\pi L}{\lambda},
\label{prob}
\end{equation}
where the oscillation wavelength $\lambda$ for a neutrino with 
energy $E$ is given by
\begin{equation}
 \lambda 
= \left[\frac{\pi {\mbox{ km}}}{5.07}\right] \left[\frac{10^{-15}}
{|\phi \Delta \gamma|}\right] \left[\frac{ {\mbox{MeV}}}{E}\right],
\label{wavelength}
\end{equation}
which in contrast to the wavelength for mass  induced neutrino 
oscillations in vacuum, is inversely proportional to the neutrino energy.
This energy dependence is very crucial in obtaining a good 
fit to the total rates without causing any problem 
with the SK spectrum~\cite{us}.  

\section{ANALYSIS}

We present here the results of our analysis only 
for active to active conversion since the results are 
qualitatively similar for  active to sterile one~\cite{us}. 

%%%%%%%%%%%%%%%%%%%%%%%%%%%%%%%%%%%%%%%%%%%%%%%%%%%%%%%%%%%%
\vglue -1.4cm 
\hglue -0.7cm 
\psfig{file=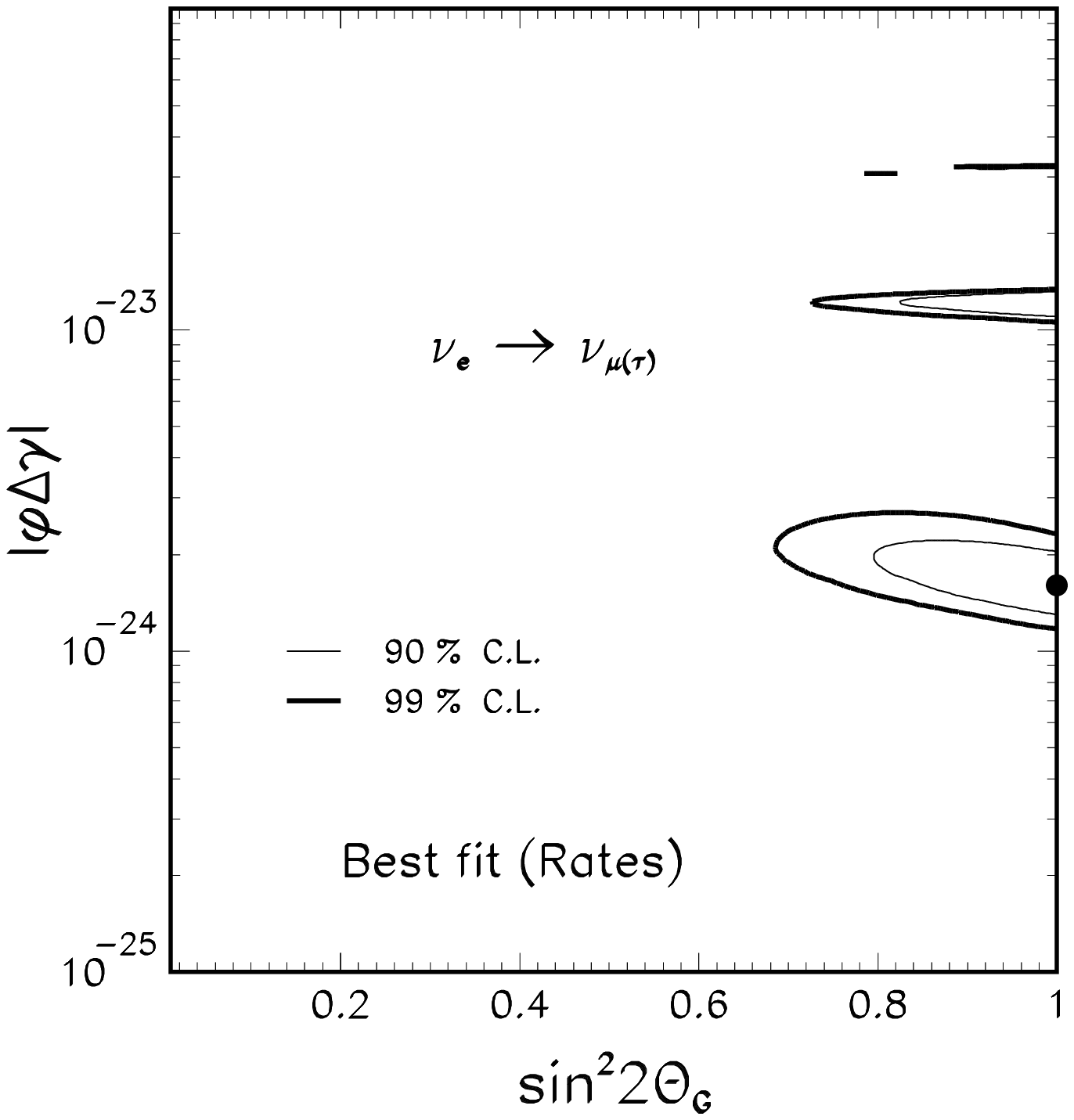,height=9.6cm,width=7.5cm}
\vglue -2.3cm 
\noindent
{\small 
Fig.1: Parameter region of $\sin^22\theta_G$ and $|\phi \Delta \gamma|$
allowed by the total rates only for $f_B=1$. 
The best fit point is indicated by the filled circle.} 
\label{fig1}
\vglue 0.3cm 
%%%%%%%%%%%%%%%%%%%%%%%%%%%%%%%%%%%%%%%%%%%%%%%%%%%%%%%%%%%%

We performed the same statistical analysis as we did 
for three flavor vaccum oscillation solution to the SNP 
in Ref.~\cite{VO3g}, which is slightly different from 
our analysis in Ref.~\cite{us}. 
We fit VEP parameters $|\phi \Delta \gamma|$ and $\theta_G$, 
and an extra normalization factor $f_B$ for the $^8$B neutrino 
flux, to the most recent experimental results coming from 
Homestake~\cite{homestake}, SAGE~\cite{sage}, 
GALLEX~\cite{gallex} and GNO~\cite{gno} combined, and 
Super-Kamiokande (SK)~\cite{sk00}.

We present in Fig.\ 1 the allowed region determined only 
by the total rates for fixed $^8$B flux. 
We then present in Fig.\ 2 the result for spectral shape 
analysis fitting only the $^8$B 
spectrum measured by SK ~\cite{sk00} in the same way as 
we did in Ref.~\cite{VO3g}.

Finally, we perform a combined fit of the rates and the spectrum 
obtaining the allowed region presented in Fig.\ 3. 
The combined allowed region is essentially the same as the one 
obtained by the rates alone.  
We also show in Fig. 4, 
the predicted spectra for the best fitted parameters, 
which are in good agreement with the data. 

%%%%%%%%%%%%%%%%%%%%%%%%%%%%%%%%%%%%%%%%%%%%%%%%%%%%%%%%%%%%
\vglue -1.5cm 
\hglue -0.5cm 
\psfig{file=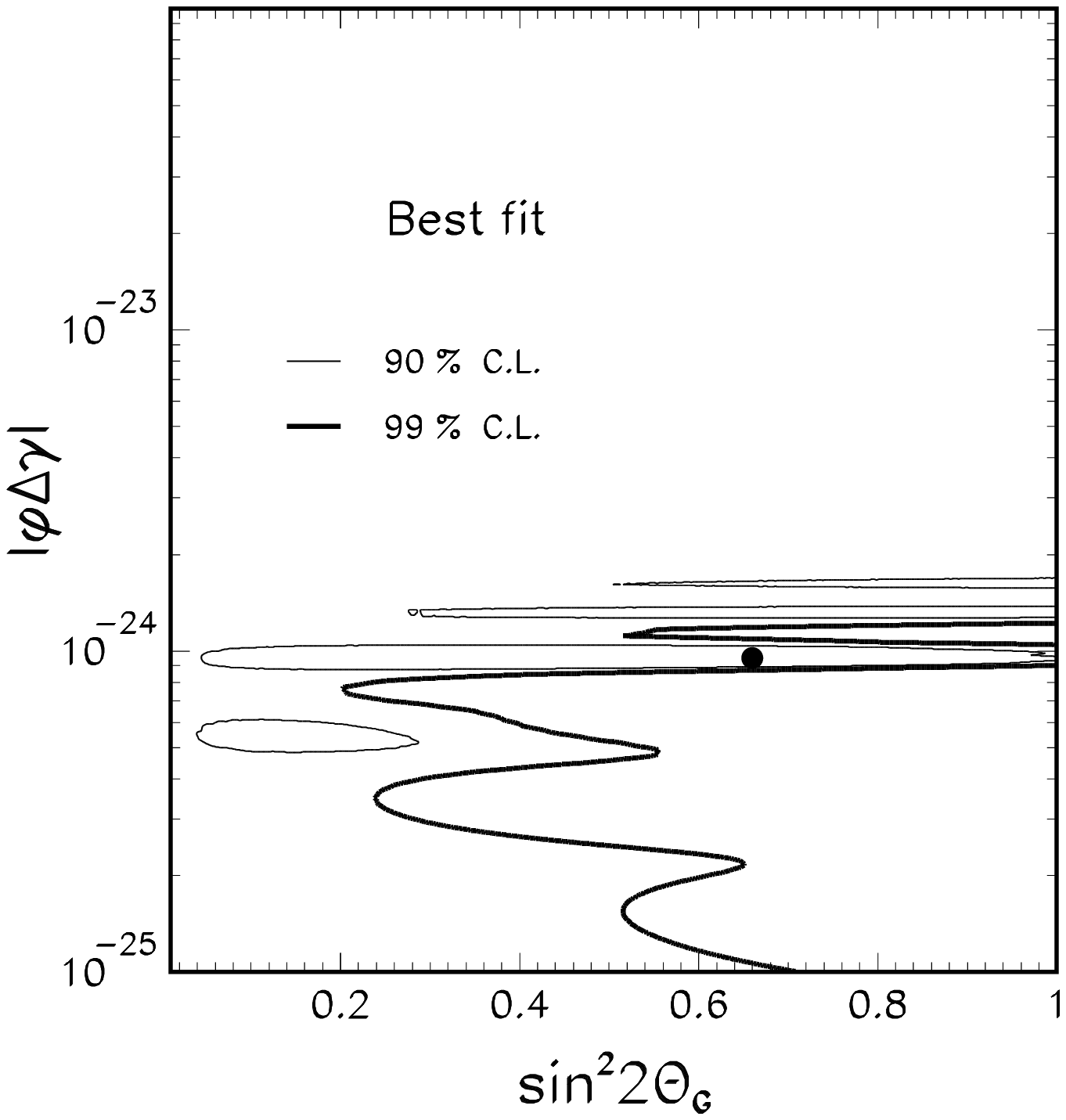,height=9.6cm,width=7.5cm}
\vglue -2.4cm 
\noindent
{\small Fig. 2: Same as in Fig. 1 but for SK recoil electron 
spectrum. 
It is the inner part of the contours which is 
excluded by the data.}
\vglue -1.5cm 
\label{fig2}
\hglue -0.5cm 
\psfig{file=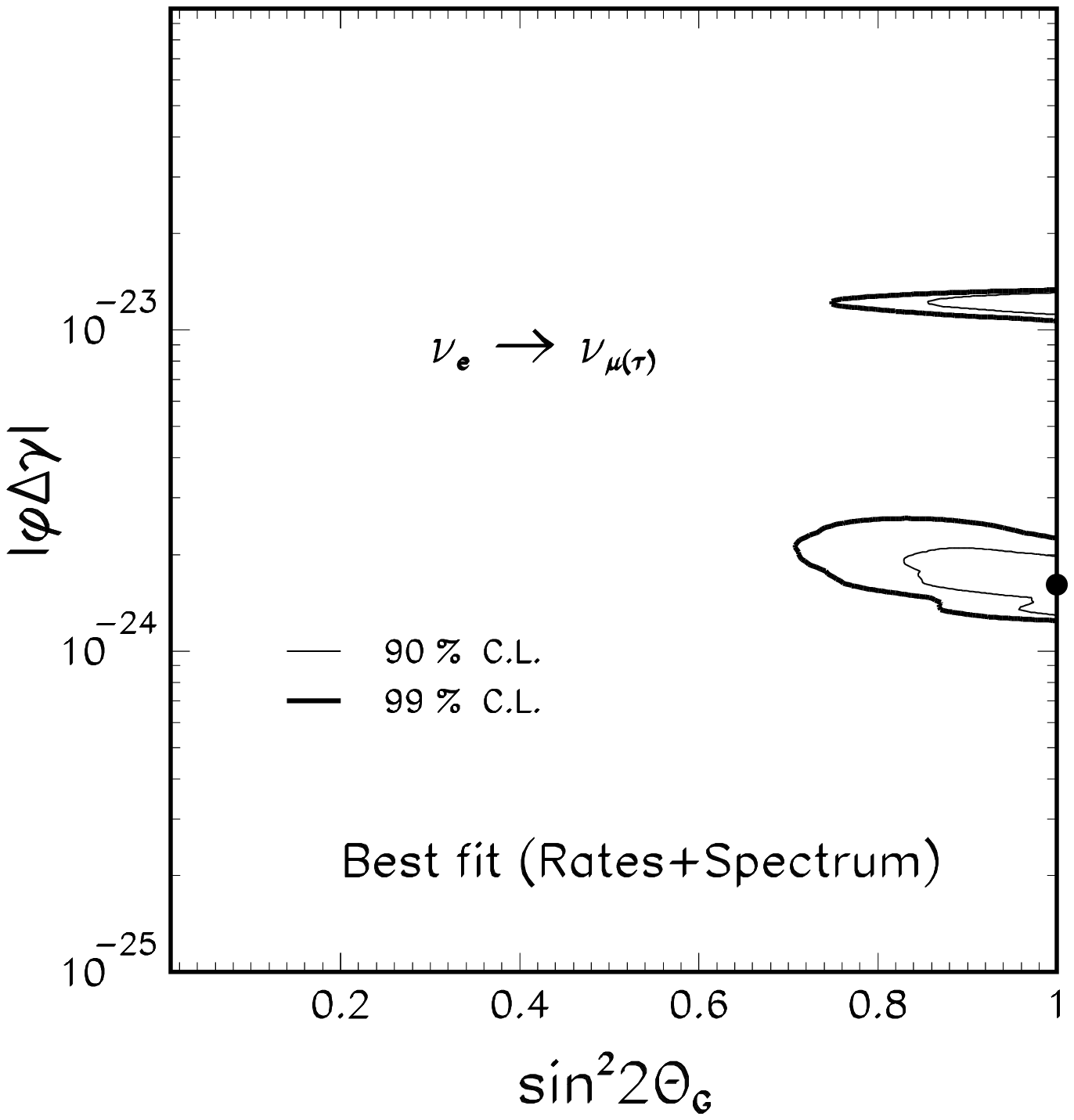,height=9.6cm,width=7.5cm}
\vglue -2.4cm 
\noindent
{\small 
Fig. 3: 
Same as in Fig. 1 but for the 
rates and SK spectrum combined. }
\vglue -0.3cm 
\label{fig3}
%%%%%%%%%%%%%%%%%%%%%%%%%%%%%%%%%%%%%%%%%%%%%%%%%%%%%%%%%%%%
\newpage 
%%%%%%%%%%%%%%%%%%%%%%%%%%%%%%%%%%%%%%%%%%%%%%%%%%%%%%%%%%
\vglue -1.6cm 
\hglue -0.3cm 
\psfig{file=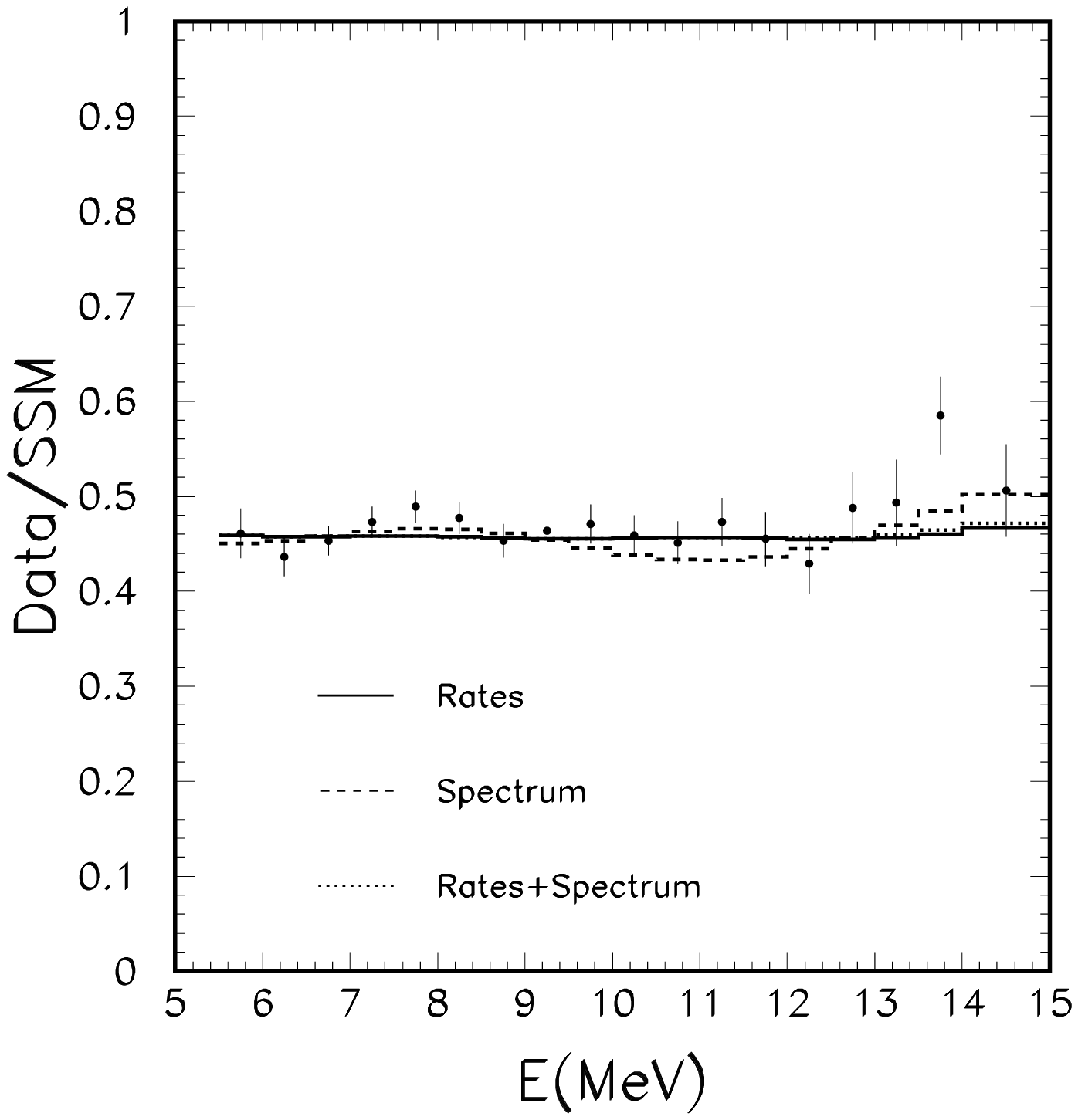,height=9.7cm,width=7.6cm}
\vglue -2.5cm 
\noindent
Fig. 4: Expected spectrum for the best fitted parameters. 
\vglue 0.2cm
%%%%%%%%%%%%%%%%%%%%%%%%%%%%%%%%%%%%%%%%%%%%%%%%%%%%%%%%%%

Best fitted parameters as well as $\chi^2_{\mbox{\scriptsize min}}$ 
can be found in 
Table I. From this table, we can conclude that the fit is quite good. 
For sterile conversion, we found somewhat worse, but acceptable, 
fit compared to the active one. For example, 
we found for the combined fit, $\chi^2_{\mbox{\scriptsize min}}$  = 19.5
for 24 D.O.F. for fixed $f_B$.  
\vglue -0.2cm
%\newpage 
\begin{table}[h]
\caption[Tab]{
\small The best fitted VEP parameters and 
values of $\chi^2_{\mbox{\scriptsize min}}$. 
Values of $|\phi \Delta \gamma|$ are shown in 
units of $10^{-24}$. 
Number of degree of freedom are from top to bottom, 
2, 1, 15, 24 and 23. $f_B$ is fixed unless it is 
indicated as ``free'' in the parenthese.
We include 6 bins for the SK zenith angle dependence 
in the combined fit. 
}
\vglue 0.2cm
\begin{center}
\begin{tabular}{ccccc}
Case  & $\sin^2 2\theta_G$ 
& $|\phi \Delta \gamma| $ & $f_B$ & 
$\chi^2_{\mbox{\scriptsize min}}$ 
\\ \hline
Rates\     &   1.0   &1.60   &  ---&  1.78  \\ 
Rates\ (free)  & 1.0   & 1.63   &  0.79&  0.39  \\ 
Spectrum   &   0.66 & 0.95   &  --- &  8.68  \\ 
Combined   &   1.0   &1.63   &  0.79&  17.65  \\ 
Combined (free)  &   1.0   &1.63   &  0.79&  16.3  \\ 
\hline
\end{tabular}
\end{center}
\label{tab1}
\vglue -1.2cm
\end{table}
\section{SUMMARY}
We have obtained a very good fit to the most recent solar 
neutrino data for the VEP induced long-wavelength neutrino 
oscillation. 

Let us finally remark that, in contrast to the usual vacuum oscillation
solution to the SNP, in this VEP 
scenario no strong seasonal effect is expected in any of the present 
or future experiments, even the ones that will be sensitive to  
$^7$Be neutrinos. See Ref.~\cite{us} for more detail. 

\vglue -0.3cm
%%%%%%%%%%%%%% Thanks
\section*{ACKNOWLEDGMENTS}
HN thanks E. Lisi for a useful discussion during the NOW2000 workshop. 
This work was supported by the Brazilian funding agencies FAPESP 
and CNPq.

\vspace{-0.1truecm}
%%%%%%%%%%%%%% References 

\end{document}